# A NONLINEAR MECHANICS-BASED VIRTUAL COILING METHOD FOR INTRACRANIAL ANEURYSM

**Seyyed Mostafa Mousavi Janbeh Sarayi (1,2), Robert J. Damiano (1,2), Palak Patel (1,2), Gary Dargush (1), Adnan H. Siddiqui (2,3), Hui Meng (1,2,3,4)**

(1) Department of Mechanical and Aerospace Engineering
University at Buffalo
Buffalo, NY, USA

(2) Canon Stroke and Vascular Research Center
University at Buffalo
Buffalo, NY, USA

(3) Department of Neurosurgery
University at Buffalo
Buffalo, NY, USA

(4) Department of Biomedical Engineering
University at Buffalo
Buffalo, NY, USA

**INTRODUCTION**

Enodvascular coils treat intracranial aneurysms (IAs) by causing them to occlude by thrombosis. Ideally, coiled IAs eventually occlude in the long-term. However, 20.8% are found incompletely occluded at treatment follow-up [1]. Computer simulations of coiling and its effect on aneurysmal flow could help clinicians predict treatment outcomes *a priori,* but it requires accurate modeling of coils and their deployment procedure. In addition to being accurate, coiling simulations must be efficient to be used as a bedside tool. To date, several virtual coiling techniques have been developed, but they lack either accuracy or efficiency. For example, finite-element-based virtual coiling methods model the mechanics of coiling and are highly accurate, at the expense of high computational cost (and thus low efficiency) [2]. Geometric-rule-based coiling techniques ignore the mechanics and therefore are computationally efficient [3], but may produce unrealistic coil deployments. In order to develop a virtual coiling method that combines accuracy and efficiency, we propose a novel virtual coiling algorithm that models coil deployment with nonlinear mechanics and nonlinear contact. Our approach is potentially more accurate than existing "simple" techniques because we model coil mechanics. It is also potentially faster than finite-element techniques because it models the most time-consuming part of these algorithms – namely contact resolution – with a novel formulation that resolves contact faster with exponential functions. Moreover, we model the coil's pre-shape as well as coil packaging into the catheter, both of which are important to model but are lacking from most existing techniques.

**METHODS**

In developing our algorithm, we considered three primary features of actual coiling procedures: coil geometry, coil packaging into the catheter, and coil deployment into the aneurysm. In Figure 1, we show these features in actual coil (left column) and in our model (right).

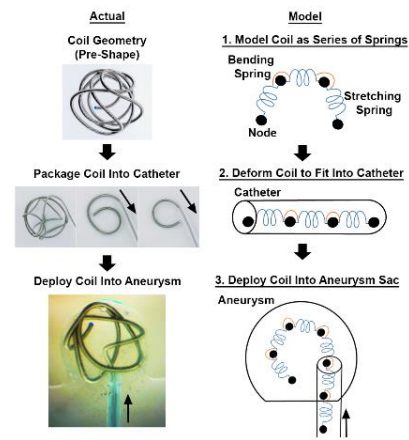

**Figure 1:** The three primary features of actual coil deployments were modeled in the virtual coiling algorithm.

**Step 1. Modeling Coil Geometry** Actual coils have a complex structure consisting of a primary wire, which is wound into a secondary spring-like structure. Moreover, in order to maximize coil distribution and clot formation in the aneurysm sac, the coils are made into a tertiary pre-shape (Figure 1, top left). To have a simplified representation of the coil structure and deployment, we modeled the coil's secondary structure as a series of equal-length linear stretching springs (with spring constants $k_s$) connected at nodes (Figure 1, right column). The length of each spring ($l_0$) was set to the diameter ($d$) of the coil wire. To account for bending, neighboring spring elements were connected to each other by linear bending springs (with spring constant $k_b$) (Figure 1, right column). Modeling a coil's geometry and mechanics as springs is more efficient than finite-element approaches, which typically model them with higher-order beam element formulations [2].

**Step 2. Deforming the Coil into the Catheter** To accurately capture the mechanics of coil deployment it is important to take into account pre-packaging of the coil into the catheter, since the straightening of the coil from its pre-shape would store strain energy, which will eventually affect the deployment and final configuration of the coil. The actual physical process of packaging the coil into the catheter is depicted in Figure 1 (left middle). To deform the virtual coil into the catheter, we first determined the change in angels and orientations of neighboring spring elements required to move the coil from its pre-shape to the catheter shape. For that purpose, local coordinates were defined on each node between spring elements using Gram-Schmidt orthogonalization [4]. Then, the coil was deformed by applying the appropriate forces to the nodes. During this step, we assumed that only bending energy changes and neglect stretching energy of the coil. The total bending energy for the coil was defined by

$$W_b = \frac{1}{2} k_b \sum_j^m (\theta_j - \theta_0), \quad (1)$$

where $m$ is the number of connections, $\theta_j$ is the angle between neighboring spring elements at the current state in the simulation and $\theta_0$ is the angle between neighboring springs in the coil's reference pre-shape.

**Step 3. Coil Deployment into the Aneurysm** The final step in the coiling algorithm is deploying the coil into the IA sac (Figure 1). In the simulation, the IA sac was represented by a closed surface of tetrahedral elements. To deploy the coil, the coil was "pushed", one spring element at a time, into the IA sac, from its deformed catheter shape. The nonlinear dynamic equation of motion for the coil during deployment is given by the following nonlinear equation:

$$m\ddot{x} + c\, sign(\dot{x})(1 - e^{-|\dot{x}|/k}) = F_i + F_{c,w} + F_{c,c} + F_p. \quad (2)$$

In this equation, the second term on the left is a nonlinear damping term in which $c$ is the damping constant and $k$ is a constant. This nonlinear damping term stabilizes the coil during deployment to avoid large deformations and sharp angles between neighboring springs which would be unrealistic in real life coil deployments. The right hand side of (2) contains all of the relevant forces that act on each of the coil nodes during deployment. $F_0$ is the force used to "push" the coil into the aneurysm and $F_i$ is the coil's internal force defined by

$$F_i = \frac{\partial W}{\partial x_i}, \quad (3)$$

where $x_i$ is the position of $i$th coil node and $W$ is the total internal energy. Although negligible during coil packaging, the total stretching energy for the coil $W_s$ becomes important during deployment:

$$W = W_b + W_s, \quad (4)$$

where $W_b$ is the bending energy defined in (1) and $W_s$ is the stretching energy of the coil:

$$W_s = \frac{1}{2} k_s \sum_i^n (l_i - l_0). \quad (5)$$

Here $n$ is the number of springs and $l_i$ is the current length of spring elements in the simulation. In (2), $F_i$ also contains the energy from $F_{c,w}$ and $F_{c,c}$, which are nonlinear coil-IA wall and coil-coil contact forces. These forces are defined as

$$F_{c,w} = |F_{c,w}| N, \quad (6a)$$

$$F_{c,c} = |F_{c,c}| \frac{(\overline{x_i} - \overline{x_j})}{norm(\overline{x_i} - \overline{x_j})}, \quad (6b)$$

where $N$ is the normal vector of one of the tetrahedral segments of the aneurysm wall in contact with a coil segment, and $x_i$ and $x_j$ are the position vectors of two coil nodes involved in the contact. The magnitude of these contact forces were defined by [5]

$$|F| = \begin{cases} 0 & for\ \delta_g \geq c_0 \\ \frac{P_0}{1-e}[(1 - \delta_g/c_0)(e^{1-\delta_g/c_0})] & for\ -6c_0 < \delta_g < c_0 \\ \frac{P_0}{1-e}[7(e^7 - 1) - (6 + \delta_g/c_0)(8e^8 - 1)] & for\ -6c_0 \geq \delta_g \end{cases}, \quad (7)$$

where $|F|$ represents $|F_{c,w}|$ and $|F_{c,c}|$. In (7), $\delta_g = |x_i - x_j|$ is the distance between the two coil nodes in contact, and $c_0$ and $P_0$ are two model parameters that can be assumed to be approximately equal to the mean local pressure and two times the root-mean-square surface roughness of the aneurysm wall. Because the nonlinear contact forces increase exponentially as the contact distance decreases in (7), the computational time to resolve contact is fast. To solve (4) iteratively, we used a semi-implicit Euler scheme. The simulation ends once the coil is completely deployed into the IA.

**Preliminary Test of the Algorithm** We implemented the algorithm in Matlab and tested the code by simulating the deployment of a $3mm \times 6cm$ Stryker coil (radius $0.014605cm$) into a patient-specific internal carotid artery IA with a diameter of $3.5mm$. The pre-shape geometry of the coil was adopted from a previous study [2].

## RESULTS
Figure 2 shows the intermediate and final results of the test coil deployment. It can be seen that the coil configuration changes over time during deployment, which echoes the real coil behavior when it is pushed into an aneurysm. This demonstrates that our nonlinear mechanics-based algorithm is able to capture the dynamics of coiling.

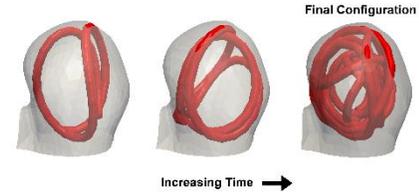

**Figure 2: Coil configuration at two intermediate time points in the test deployment simulation and the final configuration.**

## DISCUSSION
We have demonstrated a new virtual coiling algorithm that is more realistic than existing "simple" algorithms. In one preliminary example we have shown that the simulation captures the dynamic behavior of real coil deployment. This novel method is potentially faster than the most accurate finite-element algorithms. We expect greatly increased computational efficiency from: (1) modeling a coil as a series of linear and bending spring while still maintaining coil mechanics, and (2) combining nonlinear dynamics of the coil with a novel nonlinear contact mechanics formulation. In the future, we will test the robustness of our algorithm in more patient-specific IA case, compare its efficiency against existing virtual coiling methods, and experimentally validate it.